\documentclass{an}
\usepackage{graphicx}
\usepackage{times}
\Volume{00}              
\Year{0000}              
\Month{00}               

\Pagespan{001}{004}      

\begin{document}
\lhead[\thepage]{O. Almaini: Do sub-mm sources and  quasars form an evolutionary sequence?}
\rhead[Astron. Nachr./AN~{\bf XXX} (200X) X]{\thepage}
\headnote{Astron. Nachr./AN {\bf 32X} (200X) X, XXX--XXX}

\title{Do sub-mm sources and  quasars form an evolutionary sequence?}

\author{O. Almaini}
\institute{Institute for Astronomy, Royal Observatory,
Blackford Hill, Edinburgh EH9 3HJ, UK}

\date{Received {date will be inserted by the editor}; 
accepted {date will be inserted by the editor}} 

\abstract{The high redshift sub-mm sources discovered by SCUBA are
widely believed to represent the dust-enshrouded formation of massive
elliptical galaxies.  Given the strong evidence for a link between the
formation of the spheroid and the growth of the central black hole,
one might expect to see a luminous quasar at the nucleus of every
SCUBA source. Somewhat surprisingly, however, only a very small
fraction ($\sim 5$\%) are detected by Chandra with quasar luminosities.
In this paper I discuss some of the implications of these results and
discuss the accumulating evidence that sub-mm sources and quasars may
represent different stages in the evolutionary sequence of a massive
proto-spheroid.   }
\correspondence{omar@roe.ac.uk}

\maketitle

\section{Introduction}

In recent years it has become clear that essentially every massive
galaxy hosts a supermassive black hole (Kormendy \& Richstone 1995).
In particular, the remarkable relationship between the black hole mass
and the spheroidal velocity dispersion suggests a strong link
between an early epoch of quasar activity and the formation of the
spheroid (Gebhardt et al. 2000, Ferrarese \& Merritt 2000).

On a similar timescale, our understanding of the high-redshift
Universe changed dramatically with the advent of the SCUBA array at
the James Clerk Maxwell Telescope. It appears that a significant
(perhaps dominant) fraction of the star-formation in the high-redshift
Universe ($z>1$) took place in highly dust-enshrouded galaxies (Smail
et al. 1997, Hughes et al.  1998, Barger et al. 1998, Eales et
al. 1999).  These exceptionally luminous systems are the analogues of
local Ultra-Luminous Infrared Galaxies (ULIRGs).  The major difference
is that SCUBA sources appear to dominate the high-z cosmic energy
budget, while locally ULIRGS are rare and unusual events. The
discovery of the SCUBA populations was heralded by many as the
discovery of the major epoch of dust-enshrouded spheroid formation
(Lilly et al. 1999, Dunlop 2001, Granato et al. 2001).

Given the strong link suspected between the formation of the black
hole and its host spheroid, one might expect the most massive newly
forming galaxies (i.e. the SCUBA sources) to contain powerful
accreting black holes. The first Chandra observations of SCUBA
sources, however, find that only a very small fraction ($\sim 5$\%)
contain a detectable quasar (Fabian et al. 2000; Hornschemeier et
al. 2000, Severgnini  et al.  2000, Almaini et al. 2002; see Figure
1).  The implication is that the major episode of star-formation {\em
does not} coincide with the epoch of visible quasar activity.

In this paper I review the implications of these results and discuss
the possibility that sub-mm sources and quasars represent different
stages in an evolutionary sequence.

\section{The potential role of the quasar in galaxy formation} 

A number of authors have proposed models to explain the tight
black-hole/bulge mass relationship (Silk \& Rees 1998, Fabian 1999,
Archibald et al. 2002).  Although the details differ, the critical
idea is the accreting black hole grows in parallel with the bulge
until some process (possibly quasar winds) expel the gas and terminate
the star formation.

The potential importance of the black hole can be readily illustrated
when one considers the total energy generated by a quasar over its
lifetime. If we assume growth by accretion to a total mass $M$ with
the usual efficiency $\mu$ ($\sim 0.1$), folding in the typical
black-hole/bulge mass ratio of $\sim 0.1$\%, the quasar energy is
therefore $0.001\mu M_{bulge} c^2$. If we compare this to the typical
binding energy of a massive bulge, $\frac{1}{2} M_{bulge} \sigma^2$
(where $\sigma \sim 300$km s$^{-1}$), we find that less than $1$\% of
the energy generated by accretion would be sufficient to completely
destroy the host (Fabian, Wilman \& Crawford 2002). The key issue
is to determine what fraction of the quasar energy is transferred
to the host galaxy (e.g. by winds) compared to that which escapes.

A controversial possibility, therefore, is that quasars may have
played a critical role in the process of galaxy formation itself.  A
fundamental feature of these models, however, is that the bulk of the
growth of the black hole is coeval with the formation of the
spheroid. Why, then, are only $\sim 5$\% of SCUBA sources detected in
X-rays? There are at least three possible explanations, two of which
are evolutionary in nature while the other is based on a duty cycle of
AGN feeding:

\newcounter{count}
\begin{list}
{(\roman{count})}{\usecounter{count}}

\item{{\bf The duty cycle model}

In this model, the majority of the black holes within the SCUBA
sources are simply not being fueled. With star-formation rates of
$\sim 1000 M_{\odot}$yr$^{-1}$ (as required to power the most luminous
SCUBA sources) the resulting stellar winds and supernovae should
certainly generate a plentiful supply of turbulent gas, but
transfering this material to the sub-parsec scale of an accretion disk
is by no means guaranteed (Shlosman, Begelman \& Frank 1990).  A
stop-start duty cycle of fueling could therefore explain the lack of
an observable quasar at an any given epoch. This would, however, be in
stark contrast to the local counterparts of SCUBA sources (the
luminous low-z ULIRGs) where at least $50$ per cent show some evidence
evidence for an active nucleus (Sanders \& Mirabel 1996).

}

\item{{\bf Evolutionary model A: Growing black holes}

Archibald et al. (2002) have proposed a model in which the central
quasar is alive but still growing.  This is based on the hypothesis
that a black hole is likely to grow from a small seed ($\sim
100M_{\odot}$) and hence, even accreting at the Eddington limit, it
will require $\sim 5\times10^{8}$ years to reach a sufficient size to
power a quasar. At this stage the peak star-formation activity may
have ended, leading to a natural lag between the SCUBA phase and the
subsequent luminous quasar. This ``growing quasar'' model could
explain the weak levels of X-ray emission detected in many SCUBA
sources by Alexander et al. (2002) in the 2Ms observations of the
Hubble Deep Field North.

}

\item{{\bf Evolutionary model B: The Compton-thick AGN}

Perhaps the most popular explanation is that the SCUBA sources contain
accreting quasars, but these are obscured by Compton thick material
($N_H$$\gg 10^{24}$ atom cm$^{-2}$) which is later blown away as the
black hole grows (e.g. Fabian 1999).  It should be noted, however,
that a Compton-thick explanation would require almost isotropic
obscuration, rather than a standard ``Unified Scheme'' torus, since
only $\sim 5$\% of SCUBA sources are detected by Chandra.  This would
be consistent with models in which a nuclear starburst both fuels and
obscures the active nucleus (Fabian et al. 1998). Perhaps $50$ per
cent of all local AGN are Compton-thick (Maiolino et al. 1998), but
the abundance of such AGN at high redshift (and high luminosities) is
currently unclear. It is conceivable that future constraints on the
local space density of supermassive black holes may place tighter
constraints on the prevalence of Compton-thick accretion in the
Universe, particularly once the contribution of ``Compton-thin'' black
hole growth can be ascertained more accurately from the Cosmic X-ray
background.  }

\end{list}

\section{Additional observational constraints}

So far we have discussed the tight local bulge/black-hole relationship
and the observation that the majority of high-z SCUBA sources (which
are arguably proto-spheroids) are not detected as luminous X-ray sources
with Chandra. Below I outline a number of other recent observational
constraints on the links between sub-mm sources and quasars.

\subsection{Sub-mm observations of powerful high-z  quasars}

Another route to exploring the link between black-hole and spheroid
formation is to study the star-formation history of the hosts of {\em
known} active black holes.  Archibald et al. (2001) performed this
experiment for the most massive black holes by undertaking a
submillimetre study of a sample of $\simeq 50$ radio galaxies spanning
a wide range in redshift ($1<z<5$). The principal result of this study
was that the sub-mm detectability of powerful radio galaxies was found
to be a strong function of redshift, with a detection rate of $\simeq$
75\% at $z > 2.5$ compared to only $\simeq $ 15\% at $z < 2.5$. The
average sub-mm luminosity of the radio galaxies in the sample was
modeled and found to increase with redshift, with $L_{850\mu m}
\propto (1 + z)^3$, even at $z>3$.

Attempting a similar analysis with X-ray selected AGN, Page et
al. (2002) recently observed a sample of $8$ absorbed AGN over a wide
redshift range but with similar intrinsic power. Although the sample
is small, it is intriguing that only the $4$ AGN with highest redshift
($z>1.5$) were detected by SCUBA, with luminosities an order of magnitude
higher than one would expect from AGN heating.

Combined, these results provide rather convincing evidence that the
hosts of massive black holes at high redshift are very different to
the relatively passive elliptical hosts of low-redshift AGN.
Specifically, irrespective of whether the dust is heated primarily by
the UV output of young stars or by the AGN itself, it is clear that
the {\it mass} of dust in high-redshift AGN hosts is much greater at
$z > 2$.

\subsection{Sub-mm observations of typical Chandra sources}

The obvious extension of the sub-mm programmes described above is to
target more complete, flux-limited samples of AGN. It is unclear, for
example, whether radio-loud quasars have sub-mm properties which are
representative of AGN as a whole.

As a first step, we have performed a sub-mm stacking analysis at the
positions of Chandra sources in the ELAIS N2 field. We find a mean
flux of $1.25\pm0.4$mJy (Almaini et al. 2002), consistent with the
similar work of Barger et al. (2001).  Redshifts are accumulating, but
we find intriguing evidence that the stacked sub-mm flux is a strong
function of optical faintness (Figure 2). This may reflect enhanced
dust emission from obscured and/or high-z AGN, although this could be
strongly influenced by the negative K-correction which operates in the
sub-mm regime. Soon we will obtain redshifts for most of these AGN and
hence be able to study their mean sub-mm properties as a function of
both luminosity, cosmic epoch and absorbing column.

\subsection{The clustering of Chandra and SCUBA sources}

Despite the low coincidence rate, there is growing evidence that
bright SCUBA sources are strongly clustered with field AGN (Almaini et
al. 2002; see Figure 1), a result which has now been confirmed in two
further independent fields (Almaini et al. 2003, in preparation).

At first sight this would suggest that these populations, though
distinct, are tracing the same large-scale-structure and are therefore
coeval. This is puzzling, however, given that the best constraints on
the SCUBA sources suggest that the majority lie at $z>2$ (Ivison et
al. 2002, Smail et al. 2002), while the results of X-ray surveys
(e.g. Hornschemeier et al. 2001) suggest $\bar{z}\sim 1$.

I propose an alternative explanation which invokes gravitational
lensing, motivated by the particularly steep sub-mm source counts.  In
the `weak' lensing regime, for example, where the typical
magnification $\mu$ is not significantly greater than unity, one can
readily demonstrate that a population with cumulative number counts
given by a power law of index $\beta$ will be modified as follows:

\begin{equation}
N'(>S) = \mu ^{\beta -1} N(>S)
\end{equation}

Observed sub-mm number counts have a slope with $\beta \simeq 2.5$
(possibly steepening beyond 8mJy). Foreground large scale structure
(e.g. foreground groups) can readily lead to a magnification of $\mu
\simeq 1.2-1.3$.  Thus in the vicinity of such structure one would
{\em expect} an enhancement in number counts of $30-50$\% which could
easily produce a positive cross-correlation with foreground
populations (as traced by the Chandra sources). 

In conclusion, it seems likely that the appearance of large-scale
structure may simply be an artifact of gravitational lensing on a
flux-limited sub-mm survey, giving the misleading impression that
SCUBA galaxies and Chandra sources are coeval.

\section{Conclusions}

A number of independent lines of evidence now suggest that sub-mm
sources and high redshift quasars represent different stages in the
evolution of a massive proto-spheroid.  First of all, the bright SCUBA
population (arguably the progenitors of local massive elliptical
galaxies; Lilly et al. 1999) are, in general, not detected by Chandra,
despite the likelihood that these host (or will develop) black holes
in excess of $10^9 M_{\odot}$. This strongly argues for two distinct
stages of evolution, although it is unclear whether the central black hole 
is dormant, still growing or obscured by Compton-thick material.

Additional evidence comes from targeted sub-mm observations of
powerful radio galaxies, which in principle ought to be sign-posts of
massive accreting black holes irrespective of the obscuring
column. The sub-mm luminosity appears to evolve very strongly with
redshift. Similar (though more tentative) results have also emerged
from sub-mm observations of X-ray selected AGN.

Finally I note that the apparent clustering observed between SCUBA
sources and Chandra-selected AGN, which at first sight suggests that
these populations are coeval, may actually be an artifact caused by
the effects of gravitational lensing.

\begin{figure}
\resizebox{\hsize}{!}
{\includegraphics[]{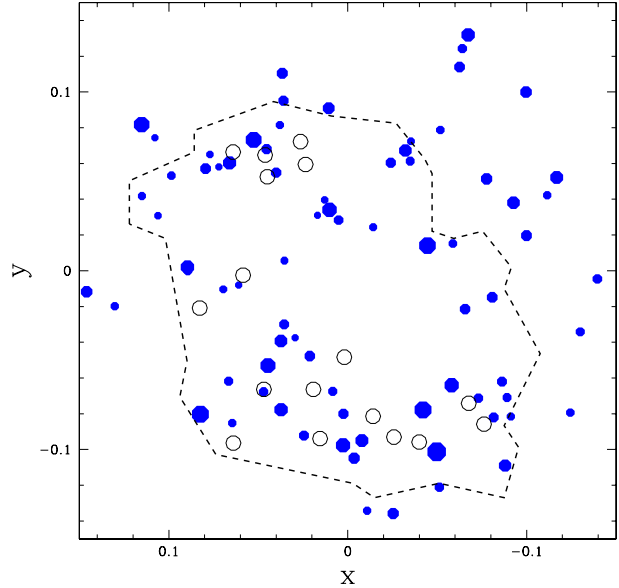}}
\caption{Illustrating the distribution of Chandra sources (filled
points) and SCUBA sources (open circles) in the ELAIS N2 field. The
size of the X-ray points are proportional to the log of their
flux. The dashed regions show the extent of the SCUBA coverage.  While
the overlap between these populations is small ($1/17$) the two
populations appear to trace the same large-scale structure (see
Almaini et al. 2002).}
\label{figlabel}
\end{figure}

\begin{figure}
\resizebox{\hsize}{!}
{\includegraphics[]{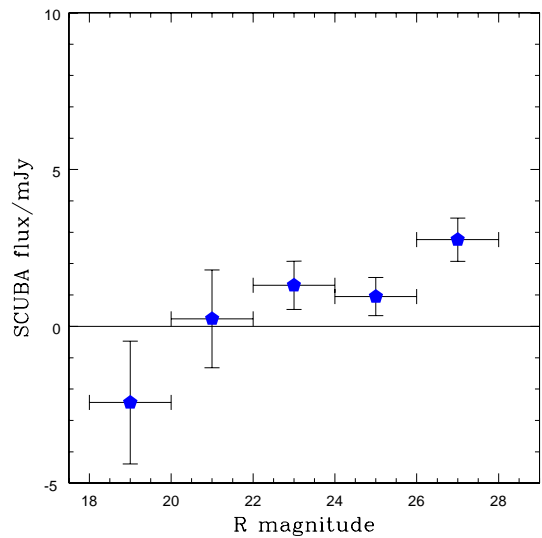}}
\caption{Stacked sub-mm flux as a function of optical magnitude for
the 55 Chandra sources lying within the SCUBA map for the ELAIS N2
field. The tentative conclusion is that obscured and/or high-z Chandra
sources may be more luminous in the sub-mm (Almaini et al. 2003, in
preparation).}
\label{figlabel}
\end{figure}

\acknowledgements

OA acknowledges the considerable support provided by the Royal Society.

\end{document}